\documentclass[journal]{IEEEtran}

\usepackage{placeins}
\usepackage{amsmath}
\usepackage{mathtools}
\usepackage{fixmath}
\usepackage{float}
\usepackage{amssymb}
\usepackage{lipsum}
\usepackage{array,multirow,graphicx}
\usepackage{rotating}
\usepackage[table]{xcolor}
\usepackage{lineno,hyperref}
\usepackage{color,soul}
\usepackage{enumitem}
\newcommand\mg[1]{{\color{black}{#1}}}
\newcommand\mh[1]{{\color{black}{#1}}}
\usepackage[ruled,longend]{algorithm2e}
\usepackage{booktabs}
\usepackage{hyperref}
\hypersetup{
	colorlinks,
	linkcolor=[rgb]{1.0, 0.0, 0.0},
	citecolor=[rgb]{1.0, 0.2, 0.2},
	urlcolor =[rgb]{0.0, 0.0, 0.8}
}
\modulolinenumbers[5]

\usepackage{pifont}% http://ctan.org/pkg/pifont
\usepackage{textcomp,array,booktabs}

\ifCLASSINFOpdf
\else
\fi
\newcommand\Euc{\mathsf{E(2)}}
\newcommand\SEuc{\mathsf{SE(2)}}
\newcommand\SOrth{\mathsf{SO(2)}}
\newcommand\reals{\mathbb{R}}
\newcommand\complexes{\mathbb{C}}
\newcommand\integers{\mathbb{Z}}

\newcommand\F{\mathcal{F}}
\newcommand\set[1]{\left\{#1\right\}}

\renewcommand\Re{\mathrm{Re}}
\newcommand\inG[2]{\rho_{#1,#2}}
\begin{document}
	\bstctlcite{IEEEexample:BSTcontrol}
	% to reduce the margin of equations and Fig.
	%\abovedisplayskip=0pt
	%\abovedisplayshortskip=0pt
	%\belowdisplayskip=0pt
	%%\belowdisplayshortskip=0pt
	%\abovecaptionskip=0pt
	%\belowcaptionskip=0pt
	
	\title{Dense Steerable Filter CNNs for Exploiting Rotational Symmetry in Histology Images}
	

	\author{Simon Graham, David Epstein
		and Nasir Rajpoot% <-this % stops a space
		\thanks{S.Graham and N.Rajpoot are with the Department of Computer Science, University of Warwick, UK.}
		\thanks{S.Graham is also with the Mathematics for Real-World Systems Centre for Doctoral Training, University of Warwick, UK.}
		\thanks{D.Epstein is with the Mathematics Institute, University of Warwick, UK.}
		}
	
	\maketitle

\begin{abstract}
Histology images are inherently symmetric under rotation, where each orientation is equally as likely to appear. However, this rotational symmetry is not widely utilised as prior knowledge in modern Convolutional Neural Networks (CNNs), resulting in \textit{data hungry} models that learn independent features at each orientation. Allowing CNNs to be rotation-equivariant removes the necessity to learn this set of transformations from the data and instead frees up model capacity, allowing more discriminative features to be learned. This reduction in the number of required parameters also reduces the risk of overfitting. In this paper, we propose Dense Steerable Filter CNNs (DSF-CNNs) that use group convolutions with multiple rotated copies of each filter in a densely connected framework. Each filter is defined as a linear combination of steerable basis filters, enabling exact rotation and decreasing the number of trainable parameters compared to standard filters. We also provide the first in-depth comparison of different rotation-equivariant CNNs for histology image analysis and demonstrate the advantage of encoding rotational symmetry into modern architectures. We show that DSF-CNNs achieve state-of-the-art performance, with significantly fewer parameters, when applied to three different tasks in the area of computational pathology: breast tumour classification, colon gland segmentation and multi-tissue nuclear segmentation. 
\end{abstract}


\IEEEpeerreviewmaketitle

\section{Introduction} \label{section:intro}

\begin{figure}[t]
		\centering
        \includegraphics[width=1.0\columnwidth]{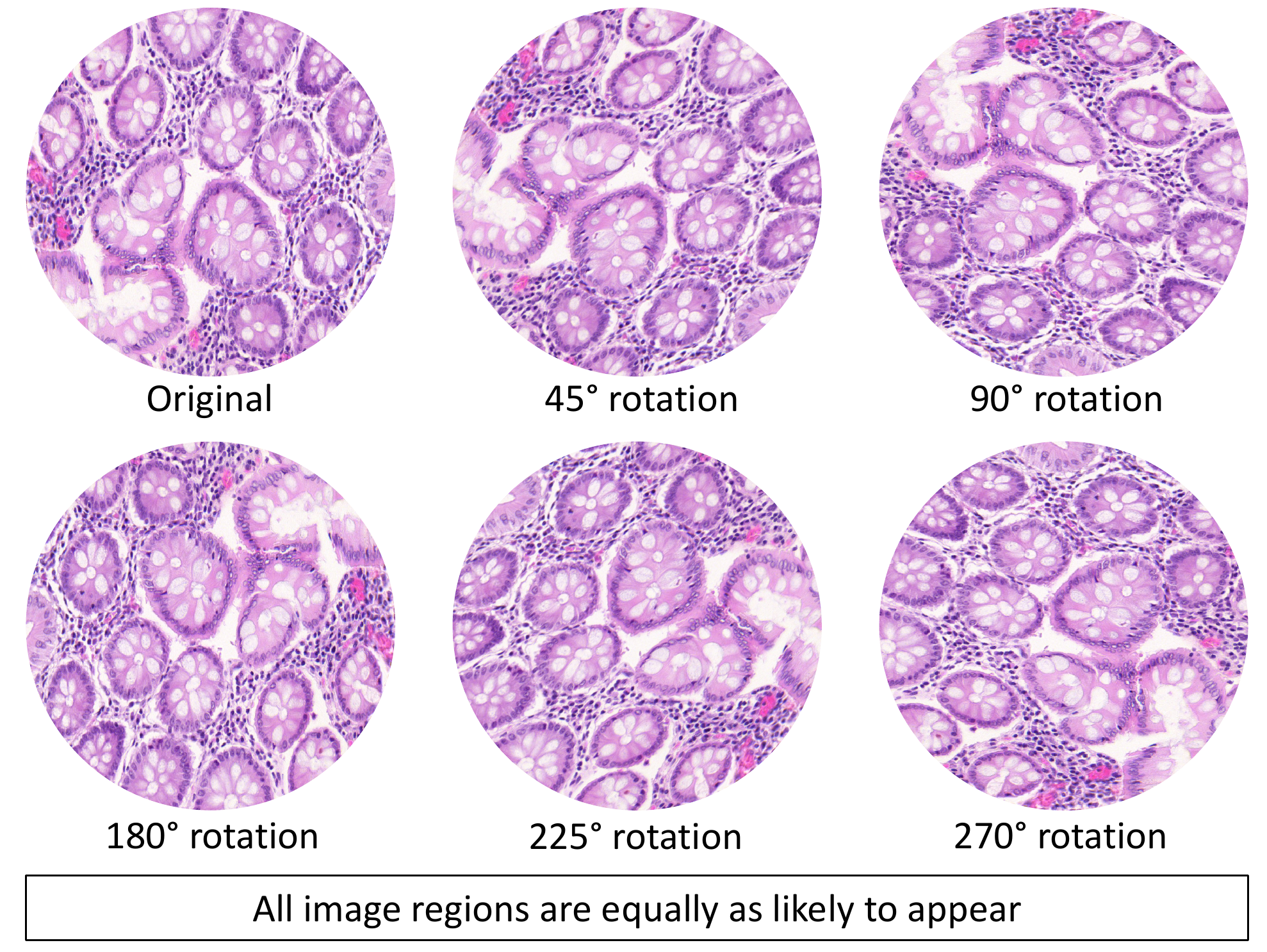}
		\caption{Cropped circular regions from a whole-slide image. Each orientation is equally as likely to appear.} 
		\label{dup1 fig:rotation}
	\end{figure}

 \IEEEPARstart{T}{he} recent advances in the analysis of Haematoxylin \& Eosin (H\&E) stained whole-slide images (WSIs) can largely be attributed to the rise of digital slide scanning \cite{snead2016validation}. In particular, Convolutional Neural Networks (CNNs) leverage the prior knowledge that images have translational symmetry and utilise a weight sharing strategy, which guarantees that a translation of the input will result in a proportional translation of the features. This property, known as \textit{translation equivariance}, is an inherent property of the CNN and removes the need to learn features at all spatial locations, significantly reducing the number of learnable parameters. In certain image analysis applications, where there is no global orientation, it is desirable to extend this property of equivariance beyond translation to also rotation. One such example is the field of computational pathology (CPath) where important image features can appear at any orientation (Fig. \ref{dup1 fig:rotation}). Therefore, we should be able to learn those features, regardless of their orientation. In the absence of rotation-equivariance, data augmentation is typically used, where multiple rotated copies of the WSI patches are usually introduced to the network during the training process. However, the augmentation strategy requires many more parameters in order to learn weights of different orientations. Instead, encoding rotational symmetry as a prior knowledge into current deep learning architectures \mg{by enforcing rotation-equivariance} requires fewer parameters and leads to an overall superior discriminative ability. \mg{Also, rotation-equivariant CNNs typically converge quicker because the network does not need to spend time learning different filter orientations.}
 
% CPath is ripe ground for the utilisation of rotation-equivariant models. Within CPath, each WSI consists of tens of thousands of various nuclei and many different histological structures, such as glands, that make up the overall \textit{tissue landscape}. It is well known that these tissue components are diagnostically informative and their appearance and organisation are linked to clinical outcome\textbf{INCLUDE REFERENCES}. Yet, in order to leverage individual histological features for downstream analysis, segmentation must be carried out as an initial step. There currently exists many segmentation models in CPath\textbf{REFERENCES}, yet they fail to exploit the existence of rotational symmetry, which results in heavily over-parameterised models with poor feature interpretability\textbf{REFERENCES}. Instead, incorporating rotation-equivariance holds great promise, where learned features are generalised over orientations. 

CPath is ripe ground for the utilisation of rotation-equivariant models, yet most models fail to incorporate this prior knowledge into the CNN architectures. Inspired by recent developments in the study of rotation-equivariant CNNs \cite{cohen2016group,weiler2018learning,marcos2017rotation,lafarge2020roto}, we propose Dense Steerable Filter based CNNs (DSF-CNNs) that integrate steerable filters \cite{freeman1991design} with group convolution \cite{cohen2016group} and a densely connected framework \cite{densenet}. \mg{Dense connectivity enables efficient gradient propagation, encourages feature re-use and consequently leads to superior performance}. Each filter is defined as a linear combination of circular harmonic basis filters, enabling exact rotation and significantly reducing the number of parameters compared to standard filters. The main contributions of this work are listed as follows:  

	\begin{itemize}
	    \item A Dense Steerable Filter CNN that achieves rotation-equivariance by integrating steerable filter group convolutions within a densely connected network.
	    \item The first thorough comparison of multiple rotation-equivariant for CPath. 
	\item We demonstrate state-of-the art performance across multiple histology image datasets with far fewer parameters. 
	\end{itemize}

\section{Related Work}

\subsection{CNNs for translation equivariance}
\mh{Images can contain numerous symmetries and therefore patterns may appear at various spatial positions and orientations. Recent methods \cite{ke2017srn} have shown that these symmetries can be detected, yet in this work we focus on how symmetries can be leveraged as a \textit{prior knowledge} to increase the performance of image recognition algorithms}. Pioneered by LeCun \textit{et al.} in 1994 \cite{lecun1998gradient}, CNNs inherently incorporate translation symmetry in images and achieve translation equivariance by re-using filters at all spatial locations. Therefore, a shift of the input leads to a proportional shift of the filter responses. This design drastically reduces the number of required parameters because features do not need to be learned independently at each location. Since the increase in computing power and the development of algorithms that assist network optimisation \cite{ioffe2015batch} CNNs have become deeper \cite{he2016deep, densenet}, leading to current state-of-the-art performance in numerous image recognition tasks \cite{imagenet_cvpr09, lin2014microsoft}. As a result of the success of deep learning, CNNs have since been widely used in CPath for various tasks including: gland segmentation \cite{graham2019mild, chen2016dcan}; nucleus segmentation \cite{graham2019hover, naylor2018segmentation, kumar2017dataset}; mitosis detection \cite{akram2018leveraging}; cancer type prediction \cite{graham2018classification} and cancer grading \cite{arvaniti2018automated, 8979298}. Yet, unlike translation, CNNs do not behave well with respect to rotation because this symmetry is not built into the network architecture. 

\subsection{Exploiting rotational symmetry}

\textbf{Rotating the data: }It is well known that histology images have no global orientation and therefore standard practice is to apply rotation augmentation to the training data \cite{tellez2019quantifying}. This improves performance, but requires many parameters and is therefore prone to overfitting. Also, there is no guarantee that CNNs trained with rotation augmentation will learn an equivariant representation and generalise to data with small rotations \cite{azulay2018deep}. To reduce the variance of predictions of multiple orientations, test-time augmentation (TTA) can be used \cite{moshkov2019test}. However, with TTA inference time scales linearly with the number of augmented copies. TI-Pooling \cite{laptev2016ti} utilises multiple rotated copies of the input in a twin network architecture, where a pooling operation over orientations is performed to find the
optimal canonical instance of the input images for training. However, like TTA, TI-Pooling is computationally expensive.

\textbf{Rotating the filters: }Cohen \& Welling \cite{cohen2016group} pioneered group equivariant CNNs ($G$-CNNs), where the convolution was generalised to share weights over additional symmetry groups beyond translation. However, they limited the filter transformation to 90$^{\circ}$ rotations and horizontal/vertical flips to ensure exact transformations on the 2D pixel grid. Veeling \textit{et al.} \cite{veeling2018rotation} showed that these $G$-CNNs can be used to improve the performance of metastasis detection in breast histology images. Furthermore, Linmans \textit{et al.} \cite{linmans2018sample} and Graham \textit{et al.} \cite{graham2019rota} extended the application of the $G$-CNNs proposed by Cohen \& Welling to pixel-based segmentation in histology images, highlighting an improved performance over conventional CNNs. The symmetries of a square grid are limited to integer translations extended by the dihedral group of order 8 (4 reflections and 4 rotations). To counter the limitation of working wih square grids in the $G$-CNN, Hoogeboom \textit{et al.} \cite{hoogeboom2018hexaconv} used hexagonal filters. However, this strategy requires images to be resampled on a hexagonal lattice, which is an additional overhead.
Instead of using exact filter rotations, Bekkers \textit{et al.} \cite{bekkers2018roto} and Lafarge \textit{et al.} \cite{lafarge2020roto} applied $G$-CNNs to several medical imaging tasks by rotating filters with bilinear interpolation. Therefore, this method was not restricted to rotations by multiples of 90$^{\circ}$, but may introduce interpolation artefacts. Oriented response networks \cite{zhou2017oriented} use active rotating filters during the convolution that explicitly encodes location and orientation information within the feature maps. 

The aforementioned methods carry forward the feature maps for each orientation throughout the network. Instead, Marcos \textit{et al.} \cite{marcos2017rotation} converted the output of multiple convolutions with rotated filter copies to a vector field by considering the magnitude and angle of the highest scoring orientation at every spatial location, leading to more compact models. To help overcome the issue of inexact filter rotation, the method only considered parameters at the centre of each filter and therefore required larger filters and consequently more parameters.

\textbf{Rotating the feature maps: }Dieleman \textit{et al.} proposed a method similar to the $G$-CNN, but instead of rotating the filters, the feature maps were rotated. This design choice has no effect on the equivariance, yet any rotation that is not a multiple of 90$^{\circ}$ may suffer from interpolation artefacts. 

\textbf{Steerable filters: }CNNs that encode rotation-equivariance are typically only equivariant to \textit{discrete} rotations. \mg{Cohen \& Welling \cite{cohen2016steerable} first proposed steerable CNNs and described a general mathematical theory that applies to both continuous and discrete groups}. To achieve full 360$^{\circ}$ equivariance, Worrall \textit{et al.} \cite{worrall2017harmonic} used the concept of steerable filters \cite{freeman1991design} and constrained the weights to be complex circular harmonics. \mg{Cheng \textit{et al.} \cite{cheng2019rotdcf} propose a rotation-equivariant CNN, named RotDCF, that decomposes filters over joint steerable bases across the space and the group geometry simultaneously}. Weiler \textit{et al.} \cite{weiler2018learning} learned steerable filters as a linear combination of atomic basis filters, which enabled exact filter rotation within $G$-CNNs. Then, these steerable filters were used within the group convolution to enable the network to be equivariant to rotation.
Weiler \& Cesa \cite{weiler2019general} then performed an extensive comparison of rotation equivariant models using steerable filters. Our method builds on the approach proposed by Weiler \textit{et al.} \cite{weiler2018learning}, by incorporating steerable filter group convolutions into a densely connected framework for superior performance.

% Group equivariant CNNs (cohen)
% bas veeling digi rotation eq
% linmans (sample efficient)
% skin lesion segmentation
% rota-net (graham)
% H-Net (worrall)
% RotEqNet (marcos)
% Bekkers
% Steerable (cohen)
% steerable filters (weiler, weiler2)
% adversarial attacks

	 % steerable filters
	 % functions to matrices
	 % cross correlation and equivariance
	 % pooling and equivariance
	 % upscampling and equivariance.
	 % resnet and skip connections	
\section{Mathematical Framework}
\label{math_framework}

In this section we present the key mathematical concepts used in our framework. We first describe images, \mg{filters} and feature maps as functions. We introduce steerable filters and describe the group-convolution ($G$-convolution) operation with these filters. This operation leads to $G$-equivariance. Below, we deal with a single filter at a time, although the method actually needs a whole filter bank to be used. We follow the method described by Weiler \textit{et al.} \cite{weiler2018learning}, but we use a slightly different formulation. We encourage readers to read both approaches.

\subsection{Images and feature maps as functions}\label{images functions}
     We model an image as a map $f:\complexes\cong\reals^2\to\reals$ with compact support\footnote{The \textit{support} of $f$ is the smallest closed subset of $\complexes$ containing $\set{z\in\complexes\mid f(z)\neq 0}$.}.
     Let $\F$ be the vector space over $\reals$ of all $f:\complexes\to\reals$, with compact support, and let $\F_\complexes$ be the vector space over $\complexes$ of all  functions $f:\complexes\to\complexes$ with compact support.
     
     We denote by $\SEuc$ the group of isometries of the plane, omitting reflections. Each element of $\SEuc$ can be written in the form $z\mapsto e^{i\theta}z + b$, where $z, b\in\complexes$ and $\theta\in\reals$.
     If $g\in\SEuc$ and $f\in\F$, we define $g.f\in\F$ by:
	 \begin{equation}\label{euc on function}
	 (g.f)(z) = f(g^{-1}(z))\text{ for }z\in\complexes.
	 \end{equation}
	 \mg{The same definition is used for $g.f:\complexes\to\complexes$ when $f\in\F_\complexes$.}
	 
	 \subsection{Steerable functions and filters:}\label{steerable functions and filters} 
	 \mg{The additive group of real numbers $\reals$ acts on $\complexes$ by rotations keeping 0 fixed. By (\ref{euc on function}), it acts linearly on $\F$ (and on $\F_\complexes$):
	 $$f^\theta(z) = f(e^{-i\theta}z) \text{ for } f\in\F,\ \theta\in\reals.$$
	 }
	 
	 We define $V(f)\subset\F_\complexes$ to be the complex vector subspace spanned by the orbit $\set{f^\theta\mid \theta\in\reals}$.
	 If $V(f)$ is a finite dimensional vector space, we say that $f$ is \textit{steerable}.
	 
	 \textbf{Theorem:} A necessary and sufficient condition for $\psi\in\F_\complexes$ to be steerable is that there should exist an integer $A\ge0$, and radial profile functions $R_k:[0,\infty)\to\complexes$ for $k \in \integers$ and $-A\le k\le A$, such that, in polar coordinates:
	 \begin{equation}\label{steerable formula complex}
	     \psi(r,\varphi) = \sum_{k=-A}^A R_k(r) e^{ik\varphi},
	 \end{equation}
	 where some or all of the radial profile functions $R_k$ may be identically zero.
	 To ensure that $\psi$ has compact support, each $R_k$ is assumed to have compact support.
	 
	 If $\psi$ satisfies (\ref{steerable formula complex}), then $V(\psi)$ is clearly finite dimensional. The reverse implication takes a bit longer to argue, but easily follows from standard theorems in Group Representation Theory\footnote{For full mathematical rigour, the theorem requires the additional hypothesis that, for each $r$, $\psi$ is a continuous function of $\varphi$. See also \cite{442980} for more technical details.}.
	 
\begin{figure}[htbp]
		\centering
        \includegraphics[width=1.0\columnwidth]{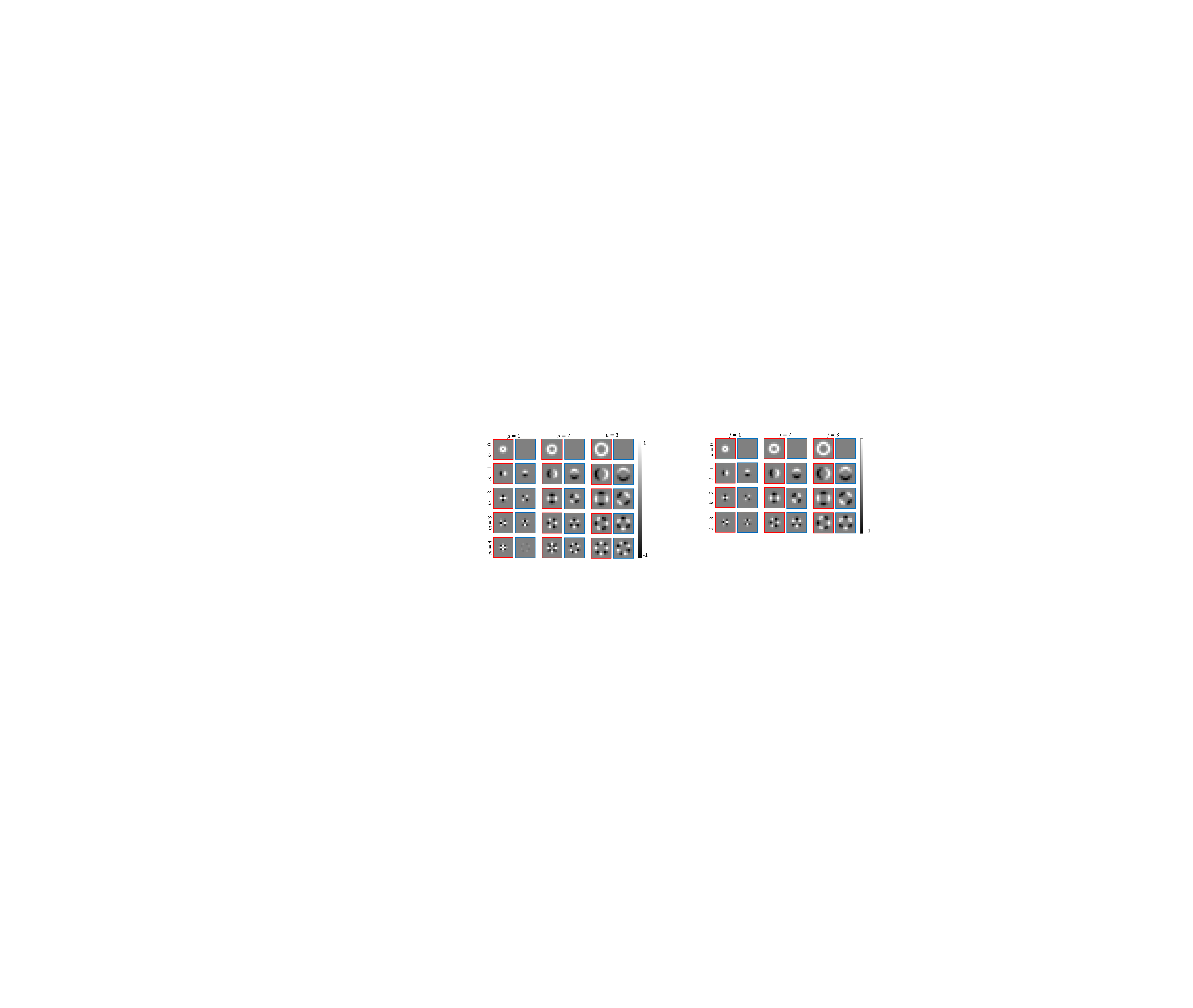}
		\caption{Example circular harmonic basis filters sampled on the 11$\times$11 square grid. Red and blue borders denote the real and imaginary parts respectively. Each pair of images comes from a single term $R_k(r) e^{ik\theta}$ in (\ref{steerable formula complex}). In this Fig., the particular radial profile functions $R_k$ are all Gaussians, as they are in our proposed model. These Gaussians have mean/mode/max at $j$. The integer $k$ specifies the frequency.} 
		\label{fig:basis filters}
	\end{figure}

	 Fig.~\ref{fig:basis filters} is a graphical representation of basis harmonic filters that appear in (\ref{steerable formula complex}).
	 
	 \textbf{Real Version:} In practice we will work with steerable real-valued filters. Since a real-valued steerable filter $\psi$ is also a complex-valued steerable filter, we can apply (\ref{steerable formula complex}) to obtain, in the same notation:
	 	 $$\psi(r,\varphi) = \Re\left(\sum_{k=-A}^A R_k(r) e^{ik\varphi}\right).$$
	 Now $\Re(z) = (z+\bar z)/2$. It follows that we can write instead (but the radial profiles change):
	 \begin{equation}\label{steerable formula real}
	     \psi(r,\varphi) = \Re\left(\sum_{k=0}^A R_k(r) e^{ik\varphi}\right)
	 \end{equation}
	 where $R_0:[0,\infty)\to\reals$ and, for $k>0$, $R_k:[0,\infty)\to\complexes$.

	 \subsection{Feature maps modelled on a group:}\label{G}
	 Following the pioneering work of Cohen and Welling \cite{cohen2016group} and of Weiler \textit{et al.} \cite{weiler2018learning}, we explain the changes to the architecture of CNNs, required to express rotation equivariance.
	 
	 \mg{We fix an integer $n>0$. We use the symbol $\inG u\theta$ to denote the euclidean transformation given by 
	 \begin{equation}\label{inG}
	 \inG{u}{\theta}(z) = e^{i\theta}z + u,
	 \end{equation}
	 where $u\in\complexes$ and $\theta = 2\pi s/n$, for some integer $s$ with $0\le s<n$.
	 Let $G\subset\Euc$ be the subgroup of all such transformations.
	 
	 Let $U$ be a group, with two subgroups $U_1$ and $U_2$. $U$ is said to be a \textit{semidirect product} of $U_1$ with $U_2$, denoted by $U_1\rtimes U_2$,
	 if there are projections $p_1:U\to U_1$ and $U\to U_2$---this means that $p_1|U_1$ and $p_2|U_2$ are both identity maps---such that $p_2$ is a homomorphism with kernel $U_1$, and $p_1\times p_2 :U\to U_1\times U_2$ is an bijection, but, in general, not an isomorphism of groups.
	 The importance of this concept in the study of equivariant CNNs was first pointed out in \cite{cohen2016group}, and there is a systematic study \cite{weiler2019general}.
	 
	 $G$ has two important subgroups, namely
	 \begin{equation}\label{Cn}
	  C_n = \{\inG 0\theta \mid \theta=2\pi s/n, 0\le s<n\},   
	 \end{equation}
	 a cyclic subgroup of order $n$ consisting of all rotations in $G$ keeping $0\in\complexes$ fixed and
	 $$T=\{\inG u0\mid u\in \complexes\} \cong \complexes,$$
	 consisting of all translations of $\complexes$. We define the group
	 \begin{equation}\label{Cn dash}
	   C_n' = \{\theta \mid \theta=2\pi s/n, 0\le s<n\},  
	 \end{equation}
	 with group law addition mod $2\pi$. Clearly, $C_n\cong C_n'$.
	 We also use $\{e\}\cong C_1$ to denote the trivial group with one element.
	 
	 The bijection
	 \begin{equation}\label{product}
	 \Pi:G\to \complexes\times C_n' \text{ defined by } \Pi(\inG u\theta)= (u,\theta)
	 \end{equation}
	 gives $G$ the semidirect product structure $G = T\rtimes C_n$. We impose on $\complexes\times C_n'$
	 a product metric that is the same as the usual Euclidean metric on $\complexes$, and is any convenient fixed metric on the finite discrete space $C_n'$. The  bijection $\Pi$
	is then used to impose a metric on $G$, so that $\Pi$ becomes an isometry. $\Pi$ does not preserve the group structure, unless $n=1$..
	
	 As a metric space $G$ is the disjoint union of the $n$ right cosets
	\begin{equation}\label{cosets}
	\complexes_\theta = T \inG 0\theta = \{\inG u\theta\mid u\in\complexes\}\subset G \text{ for } \theta\in C_n',
	\end{equation}
	such that each coset is isometric to $\complexes$.
	 
	  A \textit{$G$-feature map} is defined to be a function $f:G\to\reals$, with compact support. 
	  % It may sometimes be convenient to talk of $f$ as a feature map, omitting the $G$ prefix.
	  }

	 \mg{\subsection{$\mathcal{G}$-convolutions:}\label{H convolutions} We generalize the concept of a convolution to a $G$-convolution, that maps one $G$-feature map to another.
	  }
	  
	   We give the definition of $\mathcal{G}$-convolution, where $\mathcal{G}$\footnote{\mg{We use $\mathcal{G}$ instead of $G$ because we have reserved the name $G$ for the particular group defined in Subsection~\ref{G} and $\mathcal{G}$ denotes an arbitrary group.}} is a group with a measure $\mu_\mathcal{G}$---this means that, given $f:\mathcal{G}\to\reals$, we can form the integral denoted by
	 $\int_{g\in \mathcal{G}}f(g)\, d\mu _\mathcal{G}$ or $\int_{g\in \mathcal{G}} f(g)\,dg$. We will stick to the \textit{unimodular} case, which is general enough for all cases of interest in this paper. The word \textit{unimodular} means that we can change the dummy variable $g$ in the integral to $g^{-1}$, or $gh$ or $hg$ ($h\in \mathcal{G}$ constant), without changing the value of the integral.
	 
	 Given maps $f:\mathcal{G}\to\reals$ and $\psi:\mathcal{G}\to\reals$, we define their $\mathcal{G}$-\textit{convolution} 
	 $(f\ast_\mathcal{G} \psi):\mathcal{G}\to\reals$ by
	 \begin{align}\label{H convolution definition}
	     (f\ast_\mathcal{G} \psi)(g) & = \int_{h\in \mathcal{G}} f(gh^{-1}) \psi(h)\,dh\\
	     & = \int_{h\in \mathcal{G}} f(h)\psi(h^{-1}g)\,dh \text{ for }
	     g\in\mathcal{G}.\notag
	 \end{align}
	 The first equality is a definition, whereas the second follows by a change of variable.
	 
	 $\mathcal{G}$-convolution is automatically $\mathcal{G}$-equivariant. To see this, note that, for any $\alpha\in \mathcal{G}$,
	 \begin{align*}
	     & (\rho_\alpha(f)\ast _\mathcal{G} \psi)(g) = \int f(\alpha^{-1} g h^{-1}) \psi(h)\,dh\notag\\
	     & = (f \ast _\mathcal{G} \psi)(\alpha^{-1} g) = (\rho_\alpha(f\ast _\mathcal{G} \psi))(g).
	 \end{align*}
	 It follows that
	 \begin{equation}\label{equivariance}
	     \rho_\alpha(f)\ast_\mathcal{G} \psi = \rho_\alpha(f\ast_\mathcal{G} \psi).
	 \end{equation}
	 
	 \subsection{Hidden layer $G$-convolutions and $G$-filters}\label{G convolutions}
	 By a $G$-\textit{filter}, we mean a function $G\to\reals$. Formally this is the same as a $G$-feature map. However, in an implementation of these ideas, a $G$-feature map will turn out to be a discrete object, specified by a collection of matrices, whereas a $G$-filter retains its identity as a function. This is what enables exact rotation of a $G$-filter by an arbitrary angle.
	 
	 In order to define $G$-convolutions, we need a measure on the space $G$, as described for $\mathcal{G}$ in Subsection~\ref{H convolutions}.
	 The measure $\mu_G$ on $G$ is given by using the usual euclidean (area) measure on each $\complexes_\theta\cong \complexes$.
	 Note that $(G,\mu_G)$ is \textit{unimodular} (term defined in Subsection~\ref{H convolutions}) because rotation is measure preserving on the plane. Integration of a function $f:G\to\reals$, with respect to $\mu_G$, is carried out by first integrating each of the $n$ functions $f|\complexes_\theta\cong\complexes\to\reals$ and adding the $n$ resulting terms.

	 \mg{We now define an ``\textit{atomic steerable planar filter}'', which is not learned, but defined and does not change during training (see (\ref{atomic G})). Instead our network learns the complex coefficients used in a complex linear combination of the atomic steerable planar filters.
	 
	  For each non-negative integer $j$, we define  $\tau_j:[0,\infty)\to\reals$ to be a Gaussian, with mode at $j$, as
	 \begin{equation}\label{radial}
	 \tau_j(r) = \exp(-|r-j|^2/2\sigma^2) \text{ for } j \ge 0,\ r\ge 0.
	 \end{equation}
	 } 
	 
	 \mg{Let $j$ and $k$ be non-negative integers. By a \textit{atomic steerable planar filter}, we mean a map $\psi_{jk}:\complexes\to\complexes$ defined by
	 \begin{equation}\label{planar atomic}
	     \psi_{jk}(u) =  \tau_j(|u|) e^{ik\arg(u)}.
	 \end{equation}
	 If, in addition, $\lambda\in C_n'$, we define the \textit{atomic steerable $G$-filter} $\psi_{jk\lambda}: G\to\reals$ by
	 \begin{equation}\label{atomic G}
	     \psi_{jk\lambda}(\inG u\theta) = 
	     \begin{cases}
	     0& \text{ if } \lambda\neq\theta\\
	     \tau_j(|u|) e^{ik(\arg(u)-\theta)}& \text{ if } \lambda=\theta.
	     \end{cases}
	 \end{equation}
	 From (\ref{planar atomic}) 
	 \begin{equation}\label{2 to 3}
	     \psi_{jk\lambda}(\inG u\theta) =e^{-ik\theta}\psi_{jk}(u) \text{ if } \theta=\lambda, 
	 \end{equation}
	 which is $\psi_{jk}$ rotated by angle $\theta$.
	 
	 Any finite complex linear combination of atomic steerable $G$-filters,
	 $\sum_{j,k,\lambda} w_{jk\lambda}\psi_{jk\lambda}$, is again a steerable $G$-filter.
	 In our framework, we plan to convolve each $G$-feature map with the real part of such a sum. By (\ref{H convolution definition}) the result of such a convolution is another $G$-feature map.
	 The complex numbers $w_{jk\lambda}$ are weights in the network, determined by the network during training and each $w_{jk\lambda}$ gives rise to two real weights.
	 We will initially restrict to a single term in the finite sum, in order to keep the formulas uncluttered,
	 and then add them together.
	 
	 Let $f:G\to\reals$ be a $G$-feature map. From (\ref{H convolution definition}), we have the formula
	 \begin{equation}\label{G convolution explicit}
	 \begin{split}
	     &\left(f \ast_G \Re(w_{jk\lambda}\psi_{jk\lambda})\right)(\inG z\theta)=\\
	     &\quad\int_{\inG u\varphi\in G}
	     f(\inG u\varphi)\cdot\Re(w_{jk\lambda}\psi_{jk\lambda} (\inG v\beta))\,d\mu_G,
	     \end{split}
	 \end{equation}
	 where
	 $\inG v\beta = \inG u{\varphi}^{-1}\inG z\theta$, so that $v=e^{-i\varphi}(z-u)$ and $\beta=\theta-\varphi$.
	 From (\ref{planar atomic}) and (\ref{atomic G}),
	 \begin{equation}\label{v beta}
	     \psi_{jk\lambda}(\inG v\beta) = \begin{cases}
	     0&\text{ if } \lambda\neq \beta=\theta -\varphi\\
	     e^{-ik\varphi}\cdot\psi_{jk}(z-u)&\text{ if }\lambda=\beta=\theta - \varphi.
	     \end{cases}
	 \end{equation}
	 Writing $ f_\varphi(u) = f(\inG u\varphi)$, we obtain from (\ref{G convolution explicit}) and (\ref{v beta})
	 \begin{equation}\label{G convolution explicit 2}
	 \begin{split}
	     &\left(f \ast_G \Re(w_{jk\lambda}\psi_{jk\lambda})\right)(\inG z\theta)\\
	     &=\quad\Re\left(w_{jk\lambda}\cdot e^{-ik(\theta-\lambda)} \cdot
	     (f_{\theta-\lambda}\ast \psi_{jk}) \right)(z) \\
	    &=\quad \left(f_{\theta-\lambda} \ast \Re(w_{jk\lambda}\cdot e^{-ik(\theta-\lambda)} \psi_{jk})\right)(z).\\
	 \end{split}
	 \end{equation}
	 If we add over $\lambda\in C_n'$, then we can substitute $\varphi = \theta-\lambda$ and add over $\varphi\in C_n'$, since $\theta$ is fixed in (\ref{G convolution explicit 2}).
	 Adding over $j$, $k$ and $\varphi$, we obtain
	 \begin{equation}\label{G convolution 3}
	 \begin{split}
	     &\left(f \ast_G \Re(\sum_{jk\lambda}w_{jk\lambda}\psi_{jk\lambda})\right)(\inG z\theta)\\
	    &=\quad \sum_{jk\varphi}\left(f_\varphi \ast \Re\left(w_{jk(\theta-\varphi)}\cdot e^{-ik\varphi} \psi_{jk}\right)\right)(z)\\
	 \end{split}
	 \end{equation}
	 which recovers the same result as (10) in \cite{weiler2018learning}. We have ignored the fact that there are normally many channels ($G$-feature maps) in the domain and many channels in the range. Each pair (channel in domain, channel in base) needs its own $G$-filter, so each such pair gives rise to different weights.
	 }
	
\begin{figure*}[hbtp]
		\centering
        \includegraphics[width=1.0\textwidth]{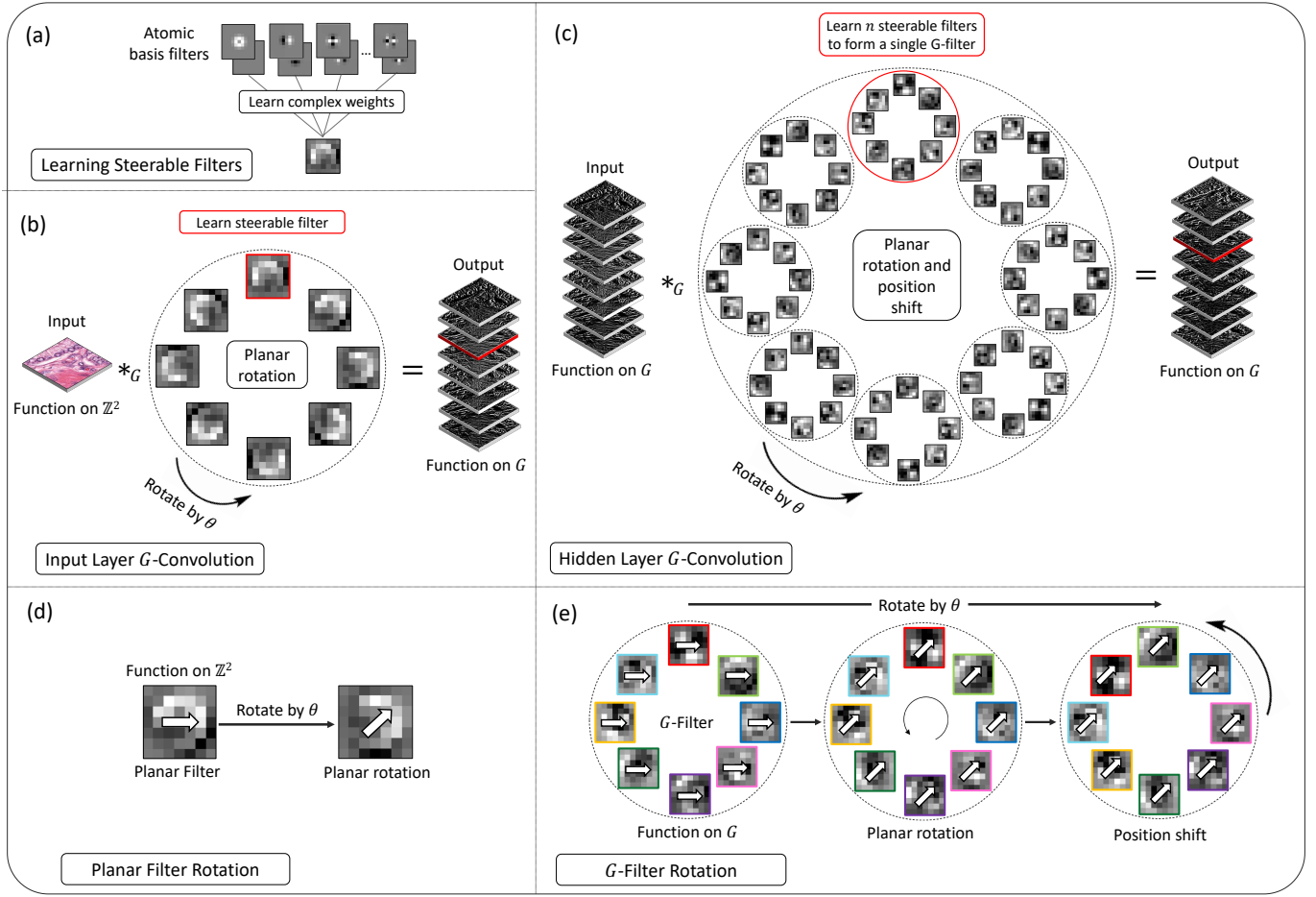}
		\caption{\mg{Overview of the two types of $G$-convolution used in our approach. with 8 filter orientations- best viewed in colour. a) Generation of steerable filters by linearly combining a series of atomic basis filters. b) Illustration  of the input layer $G$-convolution, mapping an image $f: \complexes \to\reals$ to a $G$-feature map $h:G\to\reals$. A single steerable planar filter, learned by the network, is rotated $n$ times and each rotated filter is convolved with the planar input $f$. This gives $n$ planar feature maps, which combine to give a single $G$-feature map $h$. The image $f$ is convolved with the red bordered planar filter to give the red bordered planar feature map in the stack on the right. c) Illustration of the hidden layer $G$-convolution, mapping a $G$-feature map $f:G\to\reals$ to a $G$-feature map $h:G\to\reals$. The network learns a single steerable $G$-filter, which consists of $n$ planar filters, displayed by placing them all in the same circle. Then, a single G-filter is rotated $n$ times and each rotated $G$-filter is convolved with the input $G$-feature map $f$ to generate a total of $n$ planar feature maps or a single $G$-feature map. The convolution between the input $f$ and the red circled $G$-filter gives the red bordered planar feature map on the right. d) demonstrates rotation of a planar filter, as used in the input layer $G$-convolution and e) demonstrates rotation of a $G$-filter, used in the hidden layer $G$-convolution. It can be seen from e) that $G$-filters undergo an additional position shift, in line with the group action. In both d) and e), $\theta = \frac{\pi}{4}$.
}}
		\label{dup3 fig:overview}
	\end{figure*}

	 \mg{\subsection{The input layer $G$-convolution}
	 The input to network is an image that can be thought of as a map 
	 $f:\complexes\to\reals$, which we compose with $P:G\to \complexes$ given by 
	 $P(\inG u\theta)=u$, to obtain $f\circ P:G\to \reals$.
	 By (\ref{G convolution explicit 2}), we have
	 \begin{equation*}
	 \begin{split}
	     &\left((f\circ P) \ast_G \Re(w\psi_{jk\lambda})\right)(\inG z\theta)=\\
	     &\quad\Re((w_{jk\lambda}\cdot e^{ik\lambda})\cdot e^{-ik\theta}\cdot(f\ast \psi_{jk})(z)
	 \end{split}
	 \end{equation*}
	 Since $w_{jk\lambda}$ is a complex scalar that the network has to estimate, $\lambda$ adds no new information and we dispense with it. We then sum over all terms, obtaining a simplified version of (\ref{G convolution 3}).
	 \mh{
	 \begin{equation}\label{input layer convolution}
	 \begin{split}
	     &\left((f\circ P) \ast_G \Re\left(\sum_{jk}w_{jk}\psi_{jk}\right)\right)(\inG z\theta)\\
	     &=\quad\left(f\ast \Re\left(\sum_{jk}w_{jk}\cdot e^{-ik\theta}\cdot \psi_{jk}\right)\right)(z).
	 \end{split}
	 \end{equation}
	 }
	 This gives a principled derivation of Equation~(8) in \cite{weiler2018learning}.
	 In particular, our proof of $G$-equivariance (see (\ref{equivariance})) works equally well for input layer and hidden layer $G$-convolutions. See 
	 Fig.~\ref{dup3 fig:overview}(b) for a graphical illustration of the method.
	 }
	 
	 \mg{\subsection{Sampling and the discrete case}
	 The above formulas assume that the functions involved are continuous. But a computer is a finite machine, so we need to work with discrete data, and this involves sampling.
	 
	  	 \noindent \textbf{Sampling planar steerable filters: }In the computer, a planar feature map is represented by a matrix, not by a continuous function. According to (\ref{G convolution 3}) and (\ref{input layer convolution}), we need to convolve this matrix with the real part of a complex linear combination of atomic planar filters, $\psi_{jk}$. Now $\psi_{jk}$ is a function, not a matrix---this is exactly what allows rotation of the filter through an arbitrary angle. On the other hand, convolution with a matrix requires a matrix, not a function. We therefore have to sample the atomic filters $\psi_{jk}$, and their rotations through angles $2\pi s/n$ for $0\le s<n$, at the integer points $a+ib$, where $a$ and $b$ are integers. We then perform a weighted linear combination of the sampled filters and apply (\ref{G convolution 3}) or (\ref{input layer convolution}). As the Nyquist Sampling Theorem suggests, for a fixed size of steerable filter, aliasing may occur unless one bounds the frequencies used from above. \mh{In line with Weiler \& Cesa \cite{weiler2019general}, we use frequencies up to $k = 0,2,3,2$ for $j = 0,1,2,3$ in all 7×7 steerable basis filters}. Using larger filters enables higher frequencies before aliasing, yet leads to an increase in computation time and may lead to overfitting. 
	  	 
	 \noindent \textbf{Sampling $G$-filters: }As in the case of planar convolution just discussed, our formulas need to be reinterpreted when the various component pieces of a hidden layer $G$-convolution are formulated as arrays of dimension 3 or higher, rather than as functions. For example a $G$-feature map has been defined as a function $G\to\reals$, and we need to explain how a function on the continuous group $G$ is represented in the computer by $n$ matrices.
	 
	 As shown in (\ref{cosets}), $G$ as a metric space is the disjoint union $\bigcup_{\theta\in C_n'} \complexes_\theta$ of $n$ copies of $\complexes$, with its usual euclidean metric.
	 For each $\theta\in C_n'$ (see (\ref{cosets})) we define
	  \begin{equation}\label{lattice}
	      \integers_\theta = \{\inG{a+ib}\theta \mid a,b \in\integers\}
	      \subset \complexes_\theta.
	  \end{equation}
	  Each point of $\complexes_\theta$ is within a distance $1/\sqrt2$ of some point in the lattice $\integers_\theta$.
	  It is therefore reasonable to use, as a $G$-feature map,
	 \begin{equation}
	    f: \bigcup_{\theta\in C_n'} \integers_\theta\to \reals.
	 \end{equation}
	 Analogously to the notation just before (\ref{G convolution explicit 2}), we write 
	 $f_\theta = f| \integers_\theta$.
	 The domain is infinite, but since $f$ is assumed to have compact support, we need only record the values of $f$ at a finite number of elements of $G$. In this way, a $G$-feature map is replaced by $n$ real matrices all of the same size.
	 
	 We have also defined a $G$-filter as a function $G\to\reals$. This is also sampled on $\bigcup_{\theta\in C_n'} \integers_\theta$. When learning the complex coefficients $w_{jk\lambda}$ that appear in (\ref{G convolution explicit}), the values of $j$ and $k$ are limited for the reasons just explained for the planar situation, namely to avoid aliasing and overfitting.
	 }

	 \section{Dense Steerable Filter CNN} \label{dup1 section:methods}

    \subsection{Network architecture}
	\label{network architexture}

	The main building blocks of our proposed rotation-equivariant DSF-CNN\footnote{Model code: \url{https://github.com/simongraham/dsf-cnn}} are: an input layer $G$-convolution layer; steerable filter $G$-dense-blocks and a $G$-pooling layer. Below, we build on the theoretical explanation in Section \ref{math_framework} to describe the separate components of our proposed approach. 
	
	\textbf{Input Layer $G$-convolution: }Up to the $G$-pooling operation, all convolutions within our network are steerable $G$-convolutions, as described in Section \ref{G convolutions}. Therefore, we pre-define a set of circular harmonic basis filters using (\ref{steerable formula complex}) and sample the filters on the square grid, as can be seen in Fig. \ref{fig:basis filters}. Then, we learn how to linearly combine these atomic basis filters to generate steerable filters and consider only the real part for our convolution filter, as shown in (\ref{steerable formula real}). This can be visualised in Fig. \ref{dup3 fig:overview}a. \mg{The input layer steerable $G$-convolution maps an image $f: \complexes \to\reals$ to some $G$-feature map $h:G\to\reals$. Each $G$-feature map is determined by its restriction $h_\theta$ to each coset $\complexes_\theta\cong\complexes$}. Specifically, we create $n$ rotated copies of each steerable filter and independently convolve the filters with the input to produce $n$ feature maps (or a single $G$-feature map). Planar rotation of each filter is performed using (\ref{2 to 3}) and can observed in Fig. \ref{dup3 fig:overview}d. The input layer $G$-convolution is demonstrated in Fig. \ref{dup3 fig:overview}b, where the convolution between the input and the steerable filter bordered in red produces the output also bordered in red. \mg{Now, when the input is rotated by an angle $\frac{2 \pi s}{n}$, with integers $0 \leq s < n$, and the input layer $G$-convolution is performed, the feature maps undergo a planar rotation by angle $\frac{2 \pi s}{n}$, but in addition shift $s$ positions. }
	
	\textbf{$G$-dense-blocks: }To enable efficient gradient propagation, encourage feature re-use and to improve overall performance, we use dense connectivity \cite{densenet} between $G$-convolutions in hidden layers of the network. \mg{Each hidden layer steerable $G$-convolution maps a $G$-feature map $f:G\to\reals$ to some $G$-feature map $h:G\to\reals$. We can explain this mapping in terms of the restrictions of $f$ and $h$ to cosets. Because the input to the hidden layer $G$-convolution is now a function on $G$, we must similarly ensure that our filters give a function on $G$.
	We rotate each $G$-filter to give $n$ rotated copies and perform a convolution between the input $G$-feature map $f$ and each filter orientation to produce $n$ feature maps (or a single $G$-feature map $h$).
	When rotating these $G$-filters, an additional position shift must be performed, in line with the associated group action. In Fig. \ref{dup3 fig:overview}c, $n=8$ steerable planar filters are generated as shown by the red circle, forming a single $G$-filter. This $G$-filter is convolved with the input $G$-feature map to generate the output with the red border. We can see that each $G$-filter, consists of 8 planar filters that individually rotate and shift position as the entire $G$-filter is rotated. This rotation can be seen in Fig. \ref{dup3 fig:overview}e, where the arrows show the orientation of each planar filter and the coloured borders are used to help visualise the position of each planar filter in the $G$-filter.}
	
	For each $G$-dense-block, the feature-maps of all preceding layers are concatenated to the input before performing the $G$-convolution. This increases the number of connections between layers, strengthening feature propagation. Specifically, each $G$-dense-block consists of $k$ units. Each unit contains a 7$\times$7 $G$-convolution followed by a 5$\times$5 $G$-convolution that produce 14 and 6 orientation dependent feature maps respectively. After $k$ units, the $G$-dense-block concludes by applying a final 5$\times$5 $G$-convolution. 
	
	\textbf{$G$-pooling: }At the output of the network, we transform each $G$-feature map $f$ to a planar feature map, by taking the pointwise maximum of the $n$ planar feature maps $f_\theta$ that constitute f. This operation ensures that the output of $G$-pooling is \textit{invariant} to rotation of the input.
	
	\mg{\textbf{$G$-Batch-Normalisation: }Batch normalisation (BN) involves two trainable parameters that scale and shift the normalised output. Standard BN is applied to the output of all feature maps and therefore learned BN parameters are typically different for each planar feature map in the group $G$. However, when the input is rotated, BN parameters will not transform in accordance with the input and therefore standard BN is not rotation-equivariant. Instead, after each $G$-convolution, we use a group-equivariant batch normalisation that aggregates moments per group rather than spatial feature map. This is essential to ensure rotation-equivariance throughout the network.}
	
	\textbf{Classification: }For our classification DSF-CNN, we initially perform the input layer steerable $G$-convolution followed by a hidden layer $G$-convolution. We then use 4 $G$-dense-blocks, where each block consists of 3,4,5 and 6 dense units. After every $G$-convolution layer we use a group-equivariant batch normalisation that aggregates moments per group rather than spatial feature map and ReLU non-linearity. Before every $G$-dense-block, we perform spatial max-pooling to decrease the dimensions of the feature maps. After the final $G$-dense-block, we perform $G$-pooling and then apply 3 1$\times$1 classical convolution operations to get the final output. 
	
	\textbf{Segmentation: }We extend our DSF-CNN to the task of segmentation by up-sampling feature maps after the final $G$-dense-block in the aforementioned classification CNN. Specifically, we up-sample by a factor of 2 with bilinear interpolation and then utilise a $G$-dense-block. This is repeated until the spatial dimensions of the original image are regained. From the deepest layer of the up-sampling branch, each dense-block contain 4, 3 and 2 units. In line with U-Net \cite{ronneberger2015u}, we also use skip connections to propagate information from the encoder to the decoder. After the feature maps have been up-sampled, we use a single hidden layer $G$-convolution, which is followed by $G$-pooling such that the resulting feature map is a function on $\complexes$. Finally we use 2 1$\times$1 classical convolutions to obtain the output, where we segment both the object and the contour to help separate touching instances. For nuclear segmentation, we additionally predict the eroded nuclei masks which are used as markers in marker-controlled watershed.  
	
	\section{Experiments and Results}\label{section:expandresults}
    
    \subsection{Experimental overview}
    \label{section:expoverview}
    Recently, there has been a growing number of proposed CNNs that achieve rotation-equivariance \cite{cohen2016group, weiler2018learning, marcos2017rotation, bekkers2018roto, worrall2017harmonic}, yet there is lack of comprehensive evaluation of the various methods for the analysis of histopathology images. We perform a thorough comparison of various rotation-equivariant CNNs and demonstrate the effectiveness of the proposed model. Specifically, we compare a baseline CNN with H-Nets \cite{worrall2017harmonic}, VF-CNNs \cite{marcos2017rotation}, $G$-CNNs with standard filters \cite{cohen2016group, bekkers2018roto} and $G$-CNNs with steerable filters \cite{weiler2018learning} and assess the impact of increasing the number of filter rotations in each model. For a thorough analysis, each method is applied to the task of breast tumour classification and then the best performing models are applied to the tasks of nucleus and gland segmentation. After gaining an insight into the performance of the different rotation-equivariant models, we then compare our proposed Dense Steerable Filter CNN with the state-of-the-art methods on each of the three datasets used in our experiments.
    
    		\begin{figure*}[t]
		\centering
        \includegraphics[width=1.0\textwidth]{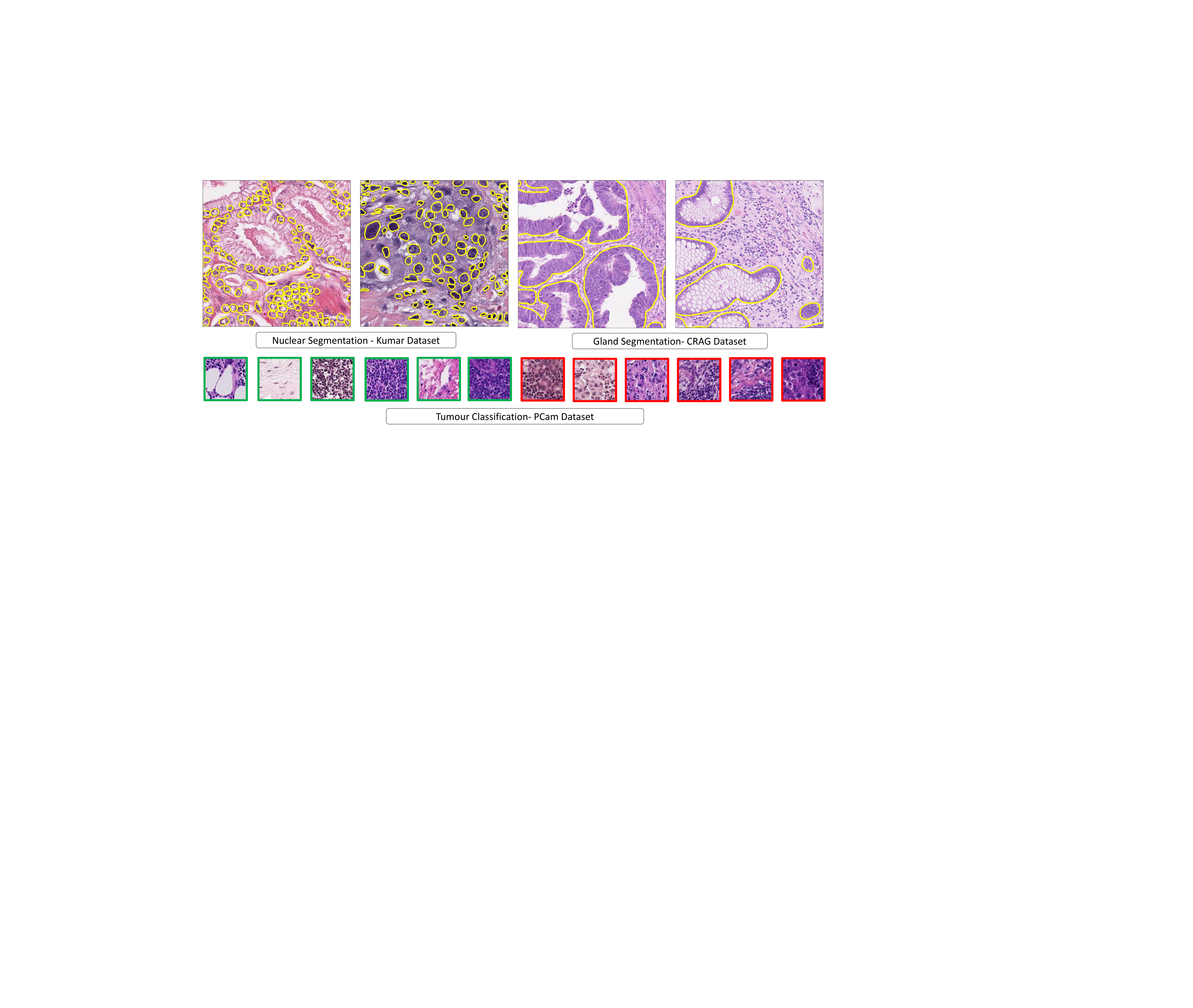}
		\caption{Image regions from the three datasets. For nuclear segmentation, gland segmentation and tumour classification, we use the Kumar \cite{kumar2017dataset}, CRAG \cite{graham2019mild} and PCam \cite{veeling2018rotation} datasets. Yellow boundaries show the pathologist annotation, while green and red borders denote non-tumour and tumour image patches.} 
		\label{fig:images}
	\end{figure*}
	
	\subsection{The three datasets} \label{section:datasets}
 	We use the following three publicly available histology image datasets:\\
 	\textbf{Breast tumour classification:} PCam \cite{veeling2018rotation} is a dataset of 327K image patches of size 96$\times$96 pixels at 10$\times$ extracted from the Camelyon16 dataset \cite{bejnordi2017diagnostic}, containing 400 H\&E stained breast WSIs. Each image patch was labelled as tumour if the central region (32$\times$32) contained at least one tumour pixel as given by the original annotation \cite{bejnordi2017diagnostic}.\\
	\textbf{Multi-tissue nucleus segmentation:} The Kumar \cite{kumar2017dataset} dataset contains 30 1,000$\times$1,000 image tiles from seven organs (6 breast, 6 liver, 6 kidney, 6 prostate, 2 bladder, 2 colon and 2 stomach) of The Cancer Genome Atlas (TCGA) database acquired at 40$\times$ magnification. Within each image, the boundary of each nucleus is fully annotated. \\
	\textbf{Colorectal gland segmentation:} The CRAG dataset \cite{graham2019mild} consists of 213 H\&E images mostly of size 1,512$\times$1,516 pixels taken from 38 WSIs acquired at 20$\times$ of colorectal adenocarcinoma (CRA) patients. It is split into 173 training images and 40 test images with different cancer grades with pixel-based gland annotation.
    
    \subsection{Evaluation metrics} \label{dup1 section:metrics}
    
    Here we describe the metrics used for evaluation. For tumour classification, we calculated the area under the receiver operating characteristic curve (AUC) to assess the binary classification performance. For gland segmentation, we employed the same quantitative measures that were used in the GlaS challenge \cite{sirinukunwattana2017gland}. These metrics consist of $F_1$, DICE and Hausdorff distance at the object level and assess the quality of instance segmentation. For nuclear segmentation, we report the binary DICE and panoptic quality (PQ). Here, the binary DICE assesses the ability of the method to distinguish nuclei from the background, whereas PQ provides insight into the quality of instance segmentation. 
    
	\subsection{Comparative analysis of rotation-equivariant models} \label{dup1 section:comparison_equiv}
	
    	\textbf{Baseline models: }For the task of breast tumour classification, we implement a baseline CNN for comparison with the aforementioned rotation-equivariant models. The model consists of a series of convolution, batch normalisation, non-linear and spatial pooling operations, which are then followed by three 1$\times$1 convolutions to obtain the final output, denoting the probability of an input patch being tumour. 
    	
        For the tasks of gland and nuclear segmentation we leverage the fully convolutional neural network architecture, which allows us to use the same model architecture, irrespective of the input size. The encoder of the baseline segmentation model uses the same architecture as the baseline classification CNN. Then a series of up-sampling and convolution operations are used to regain the spatial dimensions of the original image. In line with U-Net, we use skip connections to incorporate features from the encoder, but utilise summation as opposed to concatenation. In line with our proposed model described in Subsection~\ref{network architexture}, at the output of the network we perform segmentation of the object and the contour and additionally predict the eroded masks for nuclear segmentation. 
        
    	\textbf{Rotation-equivariant models:}\label{re_models} To assess the performance of various rotation-equivariant approaches, we modify the baseline models, but keep the fundamental architecture the same. The main difference between different models is how the filters are rotated, how many filter orientations are considered and how the convolution operation is performed.
    	
    	Aside from H-Nets, each rotation-equivariant model considers 4, 8 and 12 filter orientations. \mg{H-Nets encode full 360$^{\circ}$ equivariance within the model and therefore filters do not need to be explicitly rotated}. When applying rotation to a filter with an angle that is a multiple of $\frac{\pi}{2}$, the rotation is \textit{exact} because the output can still be represented on the square grid. However, any other rotation may give interpolation artefacts and therefore may have negative implications for rotation-equivariance. Therefore, in line with Marcos \textit{et al.} \cite{marcos2017rotation} and Lafarge \textit{et al.} \cite{lafarge2020roto}, for both the VF-CNN and standard $G$-CNN, we apply circular masking to the filters when using the groups $C_8$ and $C_{12}$. However, this masking still leads to inevitable interpolation artefacts in the centre of the filter. Steerable filters as defined by (\ref{steerable formula complex}) do not suffer from interpolation artefacts and, therefore, circular masking is not needed. 
    	
    	In all comparative experiments for rotation-equivariance, we fix each filter to be of size 7$\times$7. We used a larger filter than typically used in modern CNNs because this size ensures that we can construct a good basis set for steerable filter generation, with reasonable frequency content and reduced aliasing. 
    	
	    For fair comparison, we ensure that the number of parameters is similar between different models. For both standard and steerable $G$-CNNs, the number of parameters increases with the size of the group, \mg{if we fix the number of filters in each layer}. This is because one feature map is produced per orientation of the filter, which increases the number of required filters in the subsequent layer. To maintain the same number of parameters as the baseline CNN, we divide the number of filters in each layer of the standard $G$-CNN by $\sqrt{n}$, where $n$ is the number of orientations in the group. \mg{Steerable $G$-CNNs learn $k$ parameters (or $k/2$ complex parameters) for each filter, where typically $k < K^2$. Therefore, the number of filters in each layer of a steerable $G$-CNN should be divided by $\frac{k\sqrt{n}}{K^2}$}. Instead of carrying forward all orientations throughout the network, VF-CNNs collapse the orientation dependent feature maps to two feature maps, representing magnitude and angle. Therefore, the VF-CNN requires more filters in the next layer, but the number of parameters stays constant irrespective of the size of the group. To ensure the same number of parameters as the baseline CNN, for all group sizes we divide the number of filters in each layer of VF-CNNs by $\frac{4}{3}$. \mg{Each H-Net filter is constrained to be a complex circular harmonic, parameterised by $N$ radial terms and a single phase offset term. Also, the number of parameters is dependent on the maximum frequency $m$ of the filters. Specifically, in H-Nets frequencies in the range $[-m,m]$ are considered, equating to a total of $M=2m+1$ frequency terms. Therefore, to ensure a similar number of parameters as the standard CNN, we multiply the number of filters in each layer of a H-Net by $\frac{K^2}{M \cdot (N+1)}$.}
	    
	    In all models, we down-sample with max-pooling, but for VF-CNNs and H-Nets we use a modified pooling strategy, based on the magnitude of the feature maps. Similarly, when using both VF-CNNs amd H-Nets, we do not incorporate the angle information when using batch normalisation (BN) and non-linear activation functions; otherwise the angles may change important information about relative and global orientations. For $G$-CNNs, we use a modified BN that aggregates moments per group rather than spatial feature map. 
	    
	    To verify our implementations of the various rotation-equivariant networks, we cross-checked the performance of each model against reported benchmarks on the rotated MNIST dataset \cite{larochelle2007empirical} before applying them to the histology datasets. \mg{These results are summarised in Table \ref{dup1 table:comparative_rotmnist}.}
	 
	    \subsection{Quantitative results} \label{dup1 section:comparison_results}
	   
	    \textbf{Tumour classification: }We report comparative results of different rotation-equivariant models on the PCam dataset at the top of Table \ref{dup1 table:comparative_pcam}. \mg{We observe that H-Nets do not perform as well as the baseline CNN for the task of tumour classification. Despite this, we observe that we are able to increase the performance when incorporating higher frequency filters in the network, but the performance is still not comparable to conventional CNNs. This may suggest that constraining the filters in this way may not be optimal for detecting complex features in histology}. VF-CNNs marginally outperform the conventional CNN, where we observe that increasing the number of filter rotations leads to a slight improvement in performance. When we utilise the group convolution, with filter rotation as performed by Bekkers \textit{et al.} \cite{bekkers2018roto} and Lafarge \textit{et al.} \cite{lafarge2020roto}, we see an improved performance when using up to 8 filter orientations. This gain in performance can be attributed to incorporating our prior knowledge of rotational symmetry into the network. To ensure that we maintain a similar number of parameters, we need to reduce the number of feature maps at each layer when the size of the group is increased. This may explain the drop in performance when using 12 filter orientations. When using steerable filters, but with no filter rotation, we observe an improved performance over conventional CNNs, highlighting the benefit of learning a linear combination of basis filters, rather than standard filters. Then, as we increase the size of the group to 4 and 8 orientations we see an improvement in the performance. We also observe that using steerable filters rather than standard filters within the $G$-convolution gives a better result. 
	    
	    At the bottom of Table \ref{dup1 table:comparative_pcam} we compare the performance of our proposed DSF-CNN with the $p4m$-DenseNet \cite{veeling2018rotation}, which is the top performing method that was proposed with the introduction of the PCam dataset. This approach integrates the use of $G$-convolutions on, as proposed by Cohen \& Welling \cite{cohen2016group}, into a densely connected CNN \cite{densenet}. Here, the network uses filter rotations by multiples of 90$^{\circ}$ and also uses reflections. This is denoted by $D_4$, which is the dihedral group containing 4 rotation and 4 reflection symmetries. In addition, we compare results to the commonly used ResNet-34 \cite{he2016deep}, ResNet-50 \cite{he2016deep}, DenseNet-121 \cite{densenet} and DenseNet-169 \cite{densenet}. Despite the small amount of parameters, we observe that our method achieves the best performance with an AUC of 0.975, which is a promising improvement over the previous state-of-the-art.
	    
	    \textbf{Gland segmentation: }We compare the performance of the different rotation-equivariant models for gland segmentation on the CRAG dataset in the top part of Table \ref{dup2 table:comparative_crag}. \mg{For this experiment, when comparing different rotation-equivariant approaches, we choose to only assess the performance of conventional CNNs, standard $G$-CNNs and steerable $G$-CNNs. This is because our previous experiment on breast tumour classification indicates that $G$-CNNs are capable of achieving a superior result over competing rotation-equivariant approaches}. Similar to our observations for breast tumour classification, we see that increasing the group size within the group convolution leads to an increase in performance, but the best performance is achieved when using 8 filter orientations. For this task, using steerable filters in the group convolution led to the best performance.  
	    
	    		   \begin{table}[h]
	\begin{center}
		\caption{Tumour classification results on the PCam dataset \cite{veeling2018rotation}. Top: comparison of different rotation-equivariant models with a similar parameter budget. Bottom: comparison of proposed approach with state-of-the-art. The superscript associated with H-Net denotes the maximum frequency used.}
		\label{dup1 table:comparative_pcam}
		\setlength{\tabcolsep}{3pt} % Default value: 6pt
		\renewcommand{\arraystretch}{1.0} % Default value: 1
		\begin{tabular}{c|c|c|c}
			 \textbf{Method} & \textbf{Group}  & \textbf{Parameters}  &  \textbf{AUC}  \\
			\midrule
            CNN & \{$e$\} & 564$K$  & 0.947 \\
            H-Net$^1$ \cite{worrall2017harmonic} & $\SOrth$ & 553$K$  & 0.934 \\
            H-Net$^2$ \cite{worrall2017harmonic} & $\SOrth$ & 542$K$  & 0.939 \\
            VF-CNN \cite{marcos2017rotation} & $C_{4}$ & 556$K$ & 0.949   \\
            VF-CNN \cite{marcos2017rotation} & $C_{8}$ & 556$K$ & 0.951  \\
            VF-CNN \cite{marcos2017rotation} & $C_{12}$ & 556$K$ & 0.953   \\
            $G$-CNN \cite{cohen2016group} & $C_{4}$ & 561$K$  & 0.964   \\
            $G$-CNN \cite{bekkers2018roto,lafarge2020roto} & $C_{8}$ & 557$K$  & 0.968   \\
            $G$-CNN \cite{bekkers2018roto,lafarge2020roto} & $C_{12}$ & 557$K$  & 0.962   \\
            Steerable $G$-CNN \cite{weiler2018learning} & \{$e$\} & 553$K$  & 0.963   \\
            Steerable $G$-CNN \cite{weiler2018learning} & $C_{4}$ & 546$K$ & 0.969   \\
            Steerable $G$-CNN \cite{weiler2018learning} & $C_{8}$ & 565$K$ & 0.971   \\
            Steerable $G$-CNN \cite{weiler2018learning} & $C_{12}$ & 545$K$ & 0.969  \\
            \midrule
            ResNet-34 \cite{he2016deep}  & $\{e\}$ & 21.3$M$ &  0.942  \\
            ResNet-50 \cite{he2016deep}  & $\{e\}$ & 23.5$M$ & 0.948  \\
            DenseNet-121 \cite{densenet} & $\{e\}$ & 7.8$M$ & 0.921  \\
            DenseNet-169 \cite{densenet} & $\{e\}$ & 13.3$M$ & 0.920  \\
             $p4m$-DenseNet$^*$ \cite{veeling2018rotation}  & $D_{4}$ & 119$K$ & 0.963   \\
            DSF-CNN \textbf{(Ours)} & $C_{8}$ & 2.2$M$ & \textbf{0.975}  \\
			\bottomrule
		\end{tabular}
	\end{center}
	\end{table}

	% already defined \newcommand{\STAB}[1]{\begin{tabular}{@{}c@{}}#1\end{tabular}}
	\begin{table}[h]
		\begin{center}
			\caption{Gland segmentation results on the CRAG~\cite{graham2019mild} dataset. Top: comparison of different rotation-equivariant models with a similar parameter budget. Bottom: comparison of proposed approach with state-of-the-art.}
			\label{dup2 table:comparative_crag}
			\setlength{\tabcolsep}{2pt} % Default value: 6pt
			\renewcommand{\arraystretch}{1} % Default value: 1
				\begin{tabular}{c|c|c|c|c|c}
 					     \textbf{Method}
						 & \textbf{Group} & \textbf{Params}
						 & \textbf{Obj F$_1$} & \textbf{Obj Dice} & \textbf{Obj Haus $\downarrow$} \\
					\midrule

					%\midrule
					CNN  & \{$e$\} & 984$K$ & 0.793 &    0.809 & 246.0 \\ 
					$G$-CNN \cite{cohen2016group} & $C_4$ & 982$K$ & 0.833 & 0.856 & 170.4  \\ 
					$G$-CNN \cite{bekkers2018roto,lafarge2020roto} & $C_{8}$ & 988$K$ &  0.837 & 0.866 & 157.4 \\ 
					$G$-CNN \cite{bekkers2018roto,lafarge2020roto} & $C_{12}$ & 979$K$ &  0.818 & 0.834 & 192.2 \\ 
					Steerable $G$-CNN \cite{weiler2018learning} & \{$e$\} & 981$K$ & 0.811 & 0.848 & 175.9 \\
					Steerable $G$-CNN \cite{weiler2018learning} & $C_{4}$ & 984$K$ & 0.837 & 0.869 & 164.8 \\ 
					Steerable $G$-CNN \cite{weiler2018learning} & $C_{8}$ & 989$K$ & 0.861 & 0.888 & 139.5 \\ 
					Steerable $G$-CNN \cite{weiler2018learning} & $C_{12}$ & 976$K$ & 0.855 & 0.870 & 156.2 \\
					\midrule
					FCN8 \cite{ronneberger2015u} & \{$e$\} & 134.3$M$ &  0.796 & 0.835 & 199.5 \\
					U-Net \cite{ronneberger2015u} & \{$e$\} & 37.0$M$ & 0.827 & 0.844 & 196.9 \\ 
					MILD-Net \cite{graham2019mild} & \{$e$\} & 83.3$M$ & 0.869 & 0.883 & 146.2 \\ 
					Rota-Net \cite{graham2019rota} & $C_{4}$ & 71.3$M$ & 0.869 & 0.887 & 144.2  \\
					DSF-CNN \textbf{(Ours)} & $C_8$ & 3.7$M$ & \textbf{0.874} &  \textbf{0.891} & \textbf{138.4} \\ 
					\bottomrule
				\end{tabular}
		
		\end{center}
	\end{table}

	\newcommand{\STAB}[1]{\begin{tabular}{@{}c@{}}#1\end{tabular}}
	\begin{table}[h]
		\begin{center}
			\caption{Nuclear segmentation results on the Kumar~\cite{kumar2017dataset} dataset. Top: comparison of different rotation-equivariant models with a similar parameter budget. Bottom: comparison of proposed approach with state-of-the-art.}
			\label{dup1 table:comparative_kumar}
			\setlength{\tabcolsep}{2pt} % Default value: 6pt
			\renewcommand{\arraystretch}{1} % Default value: 1
				\begin{tabular}{c|c|c|c|c}
 					     \textbf{Method}
						 & \textbf{Group} & \textbf{Params}
						 & \textbf{B-Dice} & \textbf{PQ} \\
					\midrule

					%\midrule
					CNN  & \{$e$\} & 984$K$ & 0.767 &     0.447 \\ 
					$G$-CNN \cite{cohen2016group} & $C_4$ & 982$K$ & 0.793  & 0.490  \\ 
					$G$-CNN \cite{bekkers2018roto,lafarge2020roto} & $C_{8}$ & 988$K$ &  0.811 &  0.519 \\ 
					$G$-CNN \cite{bekkers2018roto,lafarge2020roto} & $C_{12}$ & 979$K$ &  0.814 & 0.534 \\ 
					Steerable $G$-CNN \cite{weiler2018learning} & \{$e$\} & 981$K$ & 0.791  & 0.510 \\ 
					Steerable $G$-CNN \cite{weiler2018learning} & $C_{4}$ & 984$K$ & 0.809  & 0.542 \\ 
					Steerable $G$-CNN \cite{weiler2018learning} & $C_{8}$ & 989$K$ & 0.818  & 0.543 \\ 
					Steerable $G$-CNN \cite{weiler2018learning} & $C_{12}$ & 976$K$ & 0.820  & 0.558 \\
					\midrule
					FCN8 \cite{long2015fully} & \{$e$\} & 134.3$M$ &  0.797  & 0.312 \\
					SegNet \cite{badrinarayanan2017segnet} & \{$e$\} & 29.4$M$ &  0.811 & 0.407 \\ 
					U-Net \cite{ronneberger2015u} & \{$e$\}  & 37.0$M$ &  0.758 & 0.478  \\ 
					Mask-RCNN \cite{mrcnn} & \{$e$\} & 40.1$M$ &  0.760  & 0.509 \\ 
					DIST \cite{naylor2018segmentation} & \{$e$\} & 9.2$M$ & 0.789  &  0.443 \\ 
					Micro-Net \cite{raza2019micro} & \{$e$\} & 192.6$M$ &  0.797  & 0.519  \\
					CIA-Net \cite{zhou2019cia} & \{$e$\} & 22.0$M$ &  0.818   & 0.577  \\ 
					HoVer-Net \cite{graham2019hover} & \{$e$\}  & 54.7$M$ &  \textbf{0.826}   & 0.597 \\
					DSF-CNN \textbf{(Ours)} & $C_8$  & 3.7$M$ &  \textbf{0.826}  & \textbf{0.600}   \\
					\bottomrule
				\end{tabular}
		
		\end{center}
	\end{table}
	    
	    In the bottom part of Table \ref{dup2 table:comparative_crag}, we compare our proposed approach with MILD-Net \cite{graham2019mild} and Rota-Net \cite{graham2019rota}, which are top-performing gland segmentation methods and therefore can be appropriately used for performance benchmarking. Like the $p4m$-DenseNet, Rota-Net makes use of the standard $G$-convolution, but is limited to only 90$^{\circ}$ filter rotations. In addition, we compare with FCN8 and U-Net as they are two widely used CNNs for segmentation. We observe that our DSF-CNN achieves the best performance with a fraction of the parameter budget. Notably, our model has around 20 times fewer parameters than Rota-Net and MILD-Net.
	    
	    	    	    		\begin{figure*}[t]
		\centering
        \includegraphics[width=1.0\textwidth]{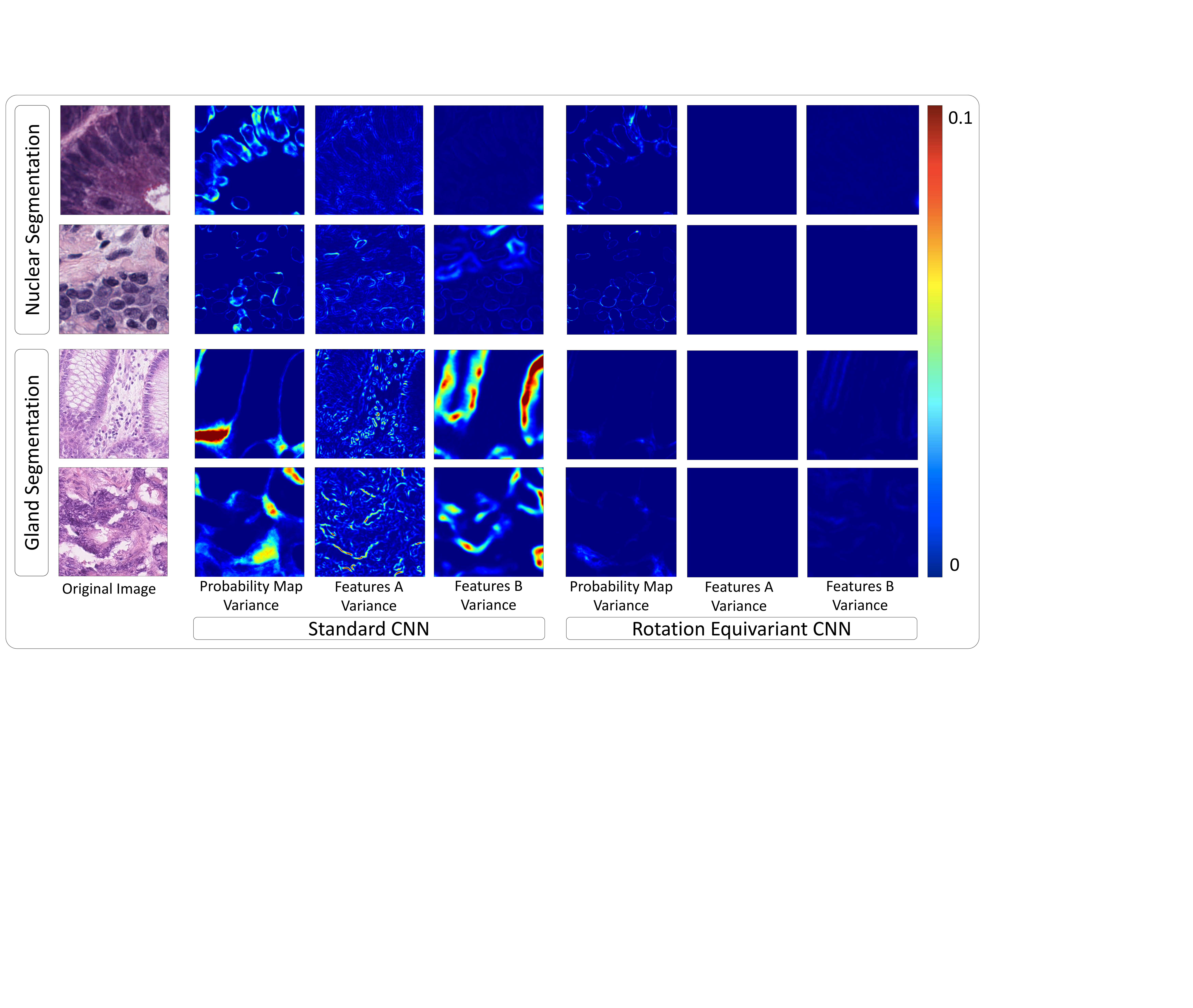}
		\caption{Variance between the predictions and features for multiple orientations of the input. The original image is rotated with steps of $\frac{\pi}{4}$ to give 8 orientations and each copy is passed through the network to enable variance calculation. Features A and B are located at the beginning and end of the network respectively. The rotation-equivariant CNN we compare with is the $C_8$ steerable $G$-CNN.}
		\label{dup1 fig:visual_results}
	\end{figure*}
	    
	    	    \begin{figure*}[t]
		\centering
        \includegraphics[width=0.99\textwidth]{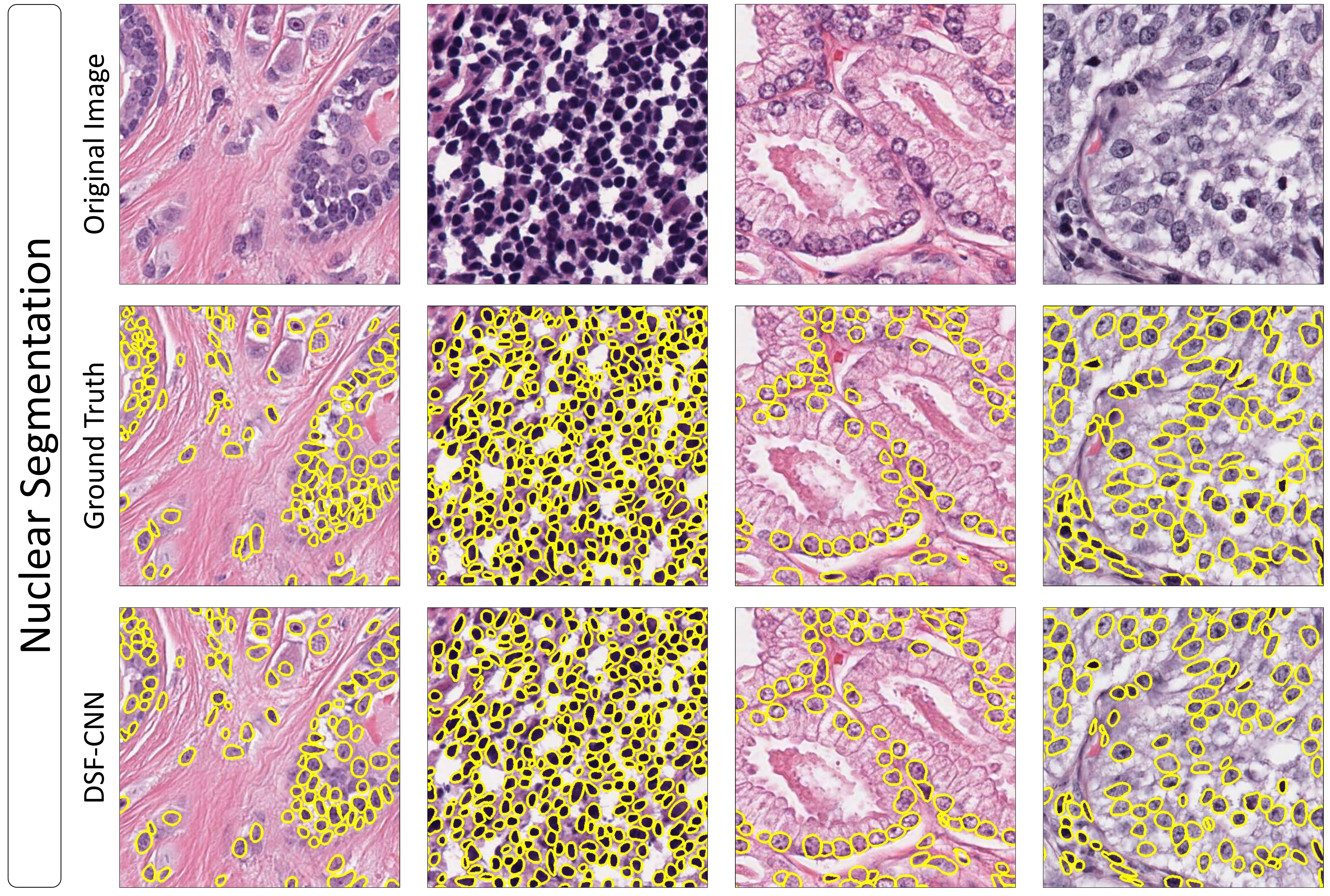}
		\caption{Visual results for nuclear segmentation on the Kumar dataset \cite{kumar2017dataset} using our proposed DSF-CNN. Yellow boundaries highlight the nuclear borders as annotated by pathologists or predicted by our algorithm.}
		\label{fig:visual_results_nuclei}
	\end{figure*}

	  \begin{figure*}[t]
		\centering
        \includegraphics[width=0.99\textwidth]{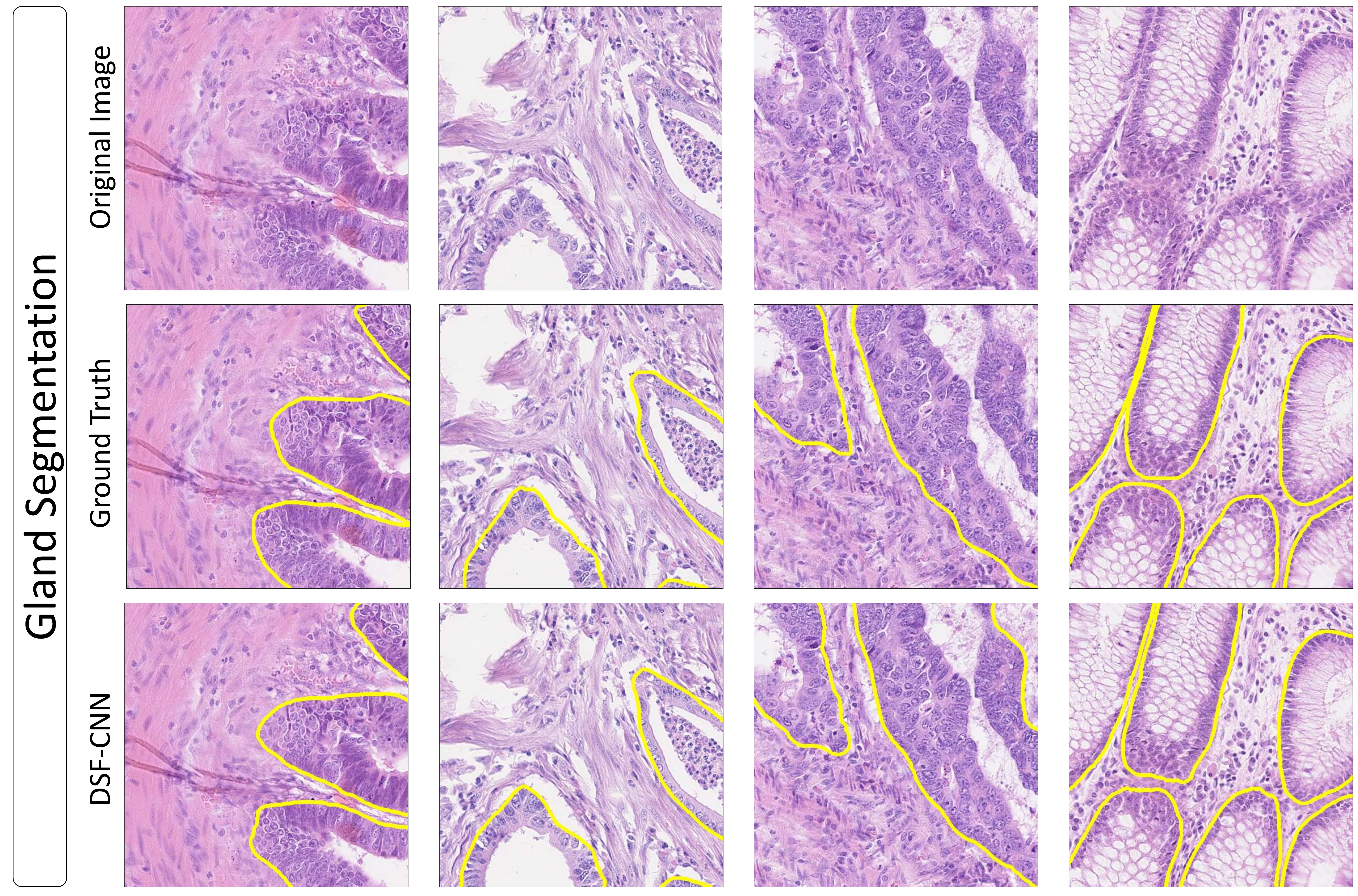}
		\caption{Visual results for gland segmentation on the CRAG dataset \cite{graham2019mild} using our proposed DSF-CNN. Yellow boundaries highlight the glandular borders as annotated by pathologists or predicted by our algorithm.}
		\label{fig:visual_results_gland}
	\end{figure*}

	    \textbf{Nuclear segmentation: }We report the comparative results of different rotation-equivariance methods for nuclear segmentation on the Kumar dataset in the top part of Table \ref{dup1 table:comparative_kumar}. \mg{Similar to above, we compare conventional CNNs with both standard and steerable $G$-CNNs}. Here, we see that all rotation-equivariant approaches show a significant improvement over standard CNNs and we see an improvement when increasing the number of filter orientations to 12 in all models. Once again, we observe that the steerable $G$-CNNs for segmentation of nuclei are superior to standard $G$-CNNs that use bilinear interpolation during filter rotation.
	    
	    We evaluate the performance of our proposed method with several state-of-the-art approaches in the bottom part of Table \ref{dup1 table:comparative_kumar}. In particular, HoVer-Net \cite{graham2019hover}, CIA-Net \cite{zhou2019cia}, Micro-Net \cite{raza2019micro} and DIST \cite{naylor2018segmentation} have been purpose-built for the task of nuclear segmentation and, therefore, provide a competitive benchmark. The proposed DSF-CNN once again achieves the best performance compared to other methods for both binary DICE and panoptic quality, on par with the state-of-the-art HoVer-Net method, while requiring a fraction of the parameter count.
	    
	    \subsection{Visual results} \label{dup1 section:comparison_equiv_viz}
	    In Fig. \ref{dup1 fig:visual_results} we visualise the features and the corresponding outputs as we rotate the input with angle increments of $\frac{\pi}{4}$ (8 in total) for both the baseline CNN and $C_8$-steerable G-CNN. Specifically, we analyse the properties of both CNNs trained for the tasks of gland and nuclear segmentation.
	    %The rotation-equivariant model that we use in this experiment is a group convolutional neural network with steerable filters at 8 orientations. 
	    To observe the feature map transformation with rotation of the input, we analyse two sets of feature maps in both CNNs: {\em Feature Map A} at the output of the 2nd convolution and {\em Feature Map B} at the output of the convolution after the final up-sampling operation. 
	    %In Fig. \ref{dup1 fig:visual_results}, we denote the feature maps at the beginning and end of the network as Features A and Features B respectively. 
	    Similarly, we observe how the output probability map transforms when the input is rotated. 
	    
	    To analyse this, we feed each image orientation into the network to obtain a set of feature maps and output probability maps. Then, after rotating features and probability maps back to their original orientation, we compute the pixel-wise variance map of the features and the output to see how they change with rotation of the input. \mg{$G$-CNN feature maps are a function on $G$ and therefore we visualise a single planar feature map within the group}. For the rotation-equivariant model, we observe that there is a near-negligible variance between the features of each input orientation. On the other hand, there is much higher variance between the features of standard CNNs after input rotation. This implies that the rotation-equivariant CNN successfully learns an equivariant feature representation. Also, there is a lower variance between the predictions of multiple input orientations for the rotation-equivariant CNN as compared to the standard CNN. Thus, the rotation-equivariant CNN behaves as expected with rotation of the input, which is a particularly desirable property when training CNNs with histology image data. \mg{It must be noted that features learned by conventional CNNs are highly complex and it is very difficult to infer the relationship between learned features and input rotation. Nonetheless, we demonstrate that rotation-equivariant CNNs have a predictable transformation with input rotation, making them more stable than conventional CNNs.}

	\subsection{Implementation and training details} \label{section:implementation}
	We implemented our framework with the open source software library TensorFlow version 1.12.0~\cite{abadi2016tensorflow} on a workstation equipped with two NVIDIA GeForce 1080 Ti GPUs. During training, data augmentation including flip, rotation, Gaussian blur and median blur was applied. For breast tumour classification, we fed the original patches of size 96$\times$96 into the network. For gland and nuclear segmentation, we used patches of size 448$\times$448 and 256$\times$256 respectively. For tumour classification, we trained our model using a batch size of 32 and then used a batch size of 8 for both gland and nucleus segmentation. We used cross-entropy loss for all tumour classification and gland segmentation models, whereas we used a combination of weighted cross-entropy and dice loss for nuclear segmentation. For all models, we trained using Adam optimisation with an initial learning rate of 10$^{-3}$, that was reduced as training progressed. The network was trained with an RGB input, normalised between 0 and 1.

\section{Discussion and Conclusions}

Conventional CNNs do not behave as expected with rotation of the input, which is a particularly undesirable property in the field of computational pathology, where important features in histology images can appear at any orientation. Instead, rotation-equivariant CNNs build this prior knowledge of rotational symmetry within the network, such that features rotate in accordance with the input without explicitly learning features at various orientations. In this paper, we proposed a densely connected steerable filter CNN that achieves state-of-the-art performance on the three datasets used in our experiments with a fraction of the parameter budget of recent top-performing models. We conducted a thorough comparative analysis of various rotation-equivariant CNNs applied to the tasks of breast tumour classification, gland segmentation and nuclear segmentation. We showed that steerable filter group convolutions gave the best quantitative results on all three tasks, where 8 filter orientations consistently gave a strong performance. We visualised features within a rotation-equivariant model to demonstrate that they rotate with the input and therefore have a higher degree of feature map interpretability. Finally, we showed that rotation-equivariant models give more stable predictions with input rotation than regular CNNs do. In future work, we will consider incorporating additional symmetries into the group convolution, such as mirror and scale symmetries. This will further increase the interpretability of feature maps and may lead to an improvement in performance and help direct future research in computational pathology.
	
\setcounter{table}{0}
\renewcommand{\thetable}{A\arabic{table}}

\FloatBarrier
\mg{
\section*{Appendix}
\setcounter{section}{0}
\addcontentsline{toc}{section}{Appendices}
\renewcommand{\thesection}{\Alph{section}}

\section{Verification of baseline models}
 In order to verify our self implemented approaches, we report the performance of each rotation-equivariant model on the rotated MNIST dataset \cite{larochelle2007empirical} in Table \ref{dup1 table:comparative_rotmnist}, which is typically used for performance benchmarking in this domain. In particular, we report the performance of a conventional CNN, H-Nets, standard $G$-CNNs, VF-CNNs and steerable $G$-CNNs. This was primarily to ensure that we were able to achieve a comparable performance with the reported results in the original papers. In our experiments all CNNs have the same base-level architecture, where we ensured that the models had the same number of layers, the same filter size and a similar number of parameters. Therefore our experiments are not only used for verification, but also to perform a fair head-to-head comparison between models. To maintain a similar number of parameters, we followed the same strategy as described in Section \ref{re_models}. In line with our experiments in the paper, for H-Net we apply spatial max-pooling based on the magnitudes, as opposed to average-pooling, which is used in the original paper.

	\begin{table}[h]
	\begin{center}
		\caption{Performance of our baseline models on rotated MNIST dataset \cite{larochelle2007empirical}. The superscript associated with H-Net denotes the maximum frequency used.}
		\label{dup1 table:comparative_rotmnist}
		\setlength{\tabcolsep}{3pt} % Default value: 6pt
		\renewcommand{\arraystretch}{1.0} % Default value: 1
		\begin{tabular}{c|c|c|c}
			 \textbf{Method} & \textbf{Group}  & \textbf{Parameters}  &  \textbf{Error}  \\
			\midrule
            CNN & \{$e$\} & 416$K$  & 2.001 \\
            H-Net$^1$ \cite{worrall2017harmonic} & $\SOrth$ & 418$K$  & 1.371   \\
            H-Net$^2$ \cite{worrall2017harmonic} & $\SOrth$ & 414$K$  & 1.352 \\
            $G$-CNN \cite{cohen2016group} & $C_{4}$ & 413$K$  & 0.976   \\
            $G$-CNN \cite{bekkers2018roto,lafarge2020roto} & $C_{8}$ & 407$K$  & 0.962   \\
            $G$-CNN \cite{bekkers2018roto,lafarge2020roto} & $C_{12}$ & 411$K$  & 0.940   \\
            VF-CNN \cite{marcos2017rotation} & $C_{8}$ & 418$K$ & 1.202   \\
            VF-CNN \cite{marcos2017rotation} & $C_{12}$ & 418$K$ & 1.172   \\
            Steerable $G$-CNN \cite{weiler2018learning} & $C_{8}$ & 416$K$ & 0.820  \\
            Steerable $G$-CNN \cite{weiler2018learning} & $C_{12}$ & 424$K$ & \textbf{0.809}  \\
			\bottomrule
		\end{tabular}
	\end{center}
	\end{table}	
	
	We observe that all rotation-equivariant CNNs achieve a greater performance than the conventional CNN, where the best performance is achieved by the $C_{12}$ steerable $G$-CNN. Interestingly, we observe a significant boost in performance for our $C_4$ $G$-CNN and H-Net implementations, compared to the originally published results. These models have the same number of layers as the original implementations, but are wider to ensure a similar number of parameters between competing models. Note, we also add 2 1$\times$1 convolutions after obtaining the invariant map (after $G$-pooling or computing the magnitude of the complex feature maps), which may have also contributed to the increase in performance. If we use the same architecture used by Weiler \textit{et al.} for the $C_{12}$ steerable $G$-CNN, then we obtain an error of 0.709, which is very close to the original result. However, this implementation uses around 3.3$M$ parameters, which is nearly 8$\times$ the amount that we use in our comparative experiments in Table \ref{dup1 table:comparative_rotmnist}.
	
		\begin{table}[h]
	\begin{center}
		\caption{Description of mathematical symbols used throughout the paper.}
		\label{dup1 table:comparative_rotmnist}
		\setlength{\tabcolsep}{3pt} % Default value: 6pt
		\renewcommand{\arraystretch}{1.0} % Default value: 1
		\begin{tabular}{c|c}
			 \textbf{Symbol} & \textbf{Description}   \\
			\midrule
            $\reals$ &  Set of real numbers \\
            $\complexes$ &  Set of complex numbers \\
            $\integers$ &  Set of integers \\ 
            $\F$ &  Real vector space of functions $\complexes\to\reals$ \\
            $\F_\complexes$ &  Complex vector space of functions
            $\complexes\to\complexes$ \\
            $\Re$ &  Real part of complex number \\
            $\Euc$ &  Euclidean group \\
            $\SEuc$ &  Special euclidean group (no reflections) \\
            $\SOrth$ &  Special orthogonal group (no reflections) \\
            \{$e$\} &   Trivial group containing only the identity on page~\pageref{Cn} \\
            $n$ & A positive integer, fixed throughout this paper\\
            $D_n$ & Dihedral group of $n$ rotations of $\complexes$, fixing 0 and flips\\
            $C_n$ & Cyclic group of $n$ rotations of $\complexes$, fixing 0\\
            $C_n'$ & $\{2\pi s/n\mid 0\le s < n\}$ group law is addition mod $2\pi$ \\
            $\mathcal{G}$ &  An arbitrary group \\
            $G$ &  Group as defined in Subsection~\ref{G}  \\
            $r$ &  radius in polar coordinates \\
            $\psi$ &  a filter \\
            $\lambda$, $\beta$, $\theta$ & usually elements of $C_n'$, sometimes arbitrary angles\\
            $R_k$ & Radial profile of atomic steerable filters \\
			\bottomrule
		\end{tabular}
	\end{center}
	\end{table}	
}
\bibliographystyle{IEEEtran}
% This version is recommended on the web, but does not work with latexmk \bibliography{IEEEabrv,ref.bib}
\bibliography{ref.bib}

\end{document}